\documentclass[12pt]{article}
\usepackage{amsfonts}
\usepackage{eurosym}
\usepackage{graphicx}
\usepackage{booktabs}
\usepackage{float}
\usepackage{subfig}
\usepackage{amssymb}
\usepackage{amsmath}
\usepackage{booktabs}
\usepackage{multirow}
\usepackage{enumerate}
\usepackage{color}
\usepackage[authoryear]{natbib}
\usepackage{hyperref}
\hypersetup{colorlinks=true, linkcolor=blue, citecolor=black, urlcolor=black}

\usepackage{setspace}
\newcommand\epigraph[2]{
\hfill{}\begin{minipage}{5.8in}{\begin{spacing}{0.9}
\noindent\textit{#1}\end{spacing}
\hfill{}{#2}}\vspace{1em}
\end{minipage}}

\begin{document}

\title{Inequality's Economic and Social Roots: \\
 the Role of Social Networks and Homophily}
\author{Matthew O. Jackson \thanks{%
Department of
Economics, Stanford University, Stanford, California 94305-6072 USA, \href{http://www.stanford.edu/\%7Ejacksonm/}{http://www.stanford.edu/$\sim$jacksonm/} ,
and an external faculty member at the Santa Fe Institute.
I gratefully
acknowledge financial support under NSF grant SES-2018554, thank Eliana La Ferrara for organizing the session, and thank Raj Chetty, Ben Golub, Mark Granovetter, Nathan Hendren, Willemien Kets, Scott Page, Anirudh Sankar, and Johannes Stroebel for helpful comments
and suggestions.
Forthcoming in {\sl Advances in Economics and Econometrics, Theory and Applications: Twelfth World Congress of the Econometric Society}, Cambridge University Press.
}}
\date{This Draft: November 2024}
\maketitle

\begin{abstract}

I discuss economic and social sources of inequality
and elaborate on the role of social networks in inequality, economic immobility, and economic inefficiencies.
The lens of social networks clarifies how the entanglement of people's information, opportunities, and behaviors
with those of their friends and family leads to persistent differences across communities, resulting
in inequality in education, employment, income, health, and wealth.
The key role of homophily in separating groups within the network is highlighted.
A network perspective's policy implications differ substantially from a narrower economic perspective that ignores social structure.
I discuss the importance of ``policy cocktails'' that include aspects that are aimed at both the economic and social forces
driving inequality.



\end{abstract}

\thispagestyle{empty}

\setcounter{page}{0} \newpage

\epigraph{Look over our vast city,  and what do we see?   On one side a very few men richer by far than it is good for a man to be, and on the other side a great mass of men and women  struggling and  worrying and wearying to get a most pitiful living.}{Henry George, October 5, 1886.}

\section{Introduction}

Given how pervasive and persistent inequality is---across places, demographics, and time---we should expect it to be driven by many forces, both independently and in concert.

For example, the economics literature has detailed how several different economic forces lead to inequality:  excess returns to capital,  monopoly rents,
market imperfections that limit people's ability to acquire human capital,
as well as
various forms of discrimination.\footnote{For some background, see  \cite*{atkinson1975,bourguignon2002inequality,bowles2002inheritance,heathcote2010unequal,banerjeed2011,benhabib2011distribution,heckman2014economics,atkinson2015inequality,alvaredo2018,benhabibb2018,smith2019capitalists,chettyj2024}.
Many additional references on specific topics appear below.}
Similarly, there are many writings on the sociology and psychology of inequality, including factors such as social stratification, institutional and geographic segregation, group competition, racism and sexism,
and cultural hegemony.\footnote{See \cite*{hurst2016social,mccall2001sources,robertsr2020,huggins2021behavioural} for some background and references.}
Even altogether, these factors miss an important perspective on inequality and key causes for many people.  These causes trace to the interplay
between social structures and economic behaviors, such as how homophily in social networks constrains
economic behaviors in labor markets, education, and other investments.

The purposes of this essay are to provide a high level overview of the many factors---both economic and social---driving inequality and how they fit together; and to highlight the interaction of social structure with economic behaviors.   Taking this wider perspective---accounting for the interaction between social structure and economic opportunities and behaviors---provides a richer understanding of how and when inequality can be systematically addressed by policies, and how policies can be better structured to take advantage of that interaction.

Humans are a heavily social species.  Social structure has enabled us to specialize and advance far beyond what any one of us
can produce \citep{henrich2015}, but with that also comes a heavy dependence upon others.
We are dependent upon our social networks\footnote{Throughout what follows, I use the term ``social network'' broadly to encompass a person's friends, neighbors, acquaintances, colleagues, and peers,  and not to refer to a social media platform.} for information, opportunities, and norms of behavior.
These dependencies tie people's fates to the fates of their friends, families, and acquaintances.

Understanding when and why this dependency leads to inequality and immobility involves recognizing that these interdependencies occur in networks that are strongly divided.
Such divides reflect the prevalence of homophily: people's tendencies to associate with others of similar demographics and backgrounds.
Given that people depend on their networks for information and opportunities, divisions in a network lead information and opportunities to stay concentrated
within parts of the society and not reach other parts.  For example, if none of a person's friends are employed in a given industry and job interviews come via referrals inside that industry, then that person is effectively shut out
of the industry.
The strong divisions and homophily that characterize social networks---that is, the tendency of people to associate with others who are similar to themselves---imply that different parts of a network can have very different information, access, and cultures.
This provides a foundation for persistent inequality, and operates through many channels including employment, but also education and other investments.

Here,  I first provide an overview of key economic\footnote{Here I artificially refer to ``economic'' as drivers that do not include
any social structure other than family, inheritance, or broad forms of discrimination; and in
which network structures are not considered.} drivers of inequality - concentration of wealth and capital and increasing returns to investments,
imperfect competition and rents to monopolies, frictions in borrowing and constraints in development of human capital, other forms of poverty traps, and discrimination.
The first two of these help explain some aspects of wealth inequality, while the others provide insights into other forms of inequality, including in education, employment and wages.
Developing a fuller understanding of inequalities, especially in these latter three areas,
requires adding social structures to help see how networks drive and sustain differences in education, employment, and wages.
Roughly, the economic drivers focus on financial and other resources, how those multiply, and how a lack of them results in lower investments and poorer outcomes.
Divides in social networks lead to three additional forces behind inequality:  unequal access to jobs and other opportunities, unequal distributions
of awareness of opportunities and information about how to take advantage of them,  and differences in norms and cultures.\footnote{\cite{coleman1988} discusses three forms of social capital that are important in the development of human capital.
There is some parallel between two of the three types of interactions here, and
two of Coleman's three types of social capital.  Inequality in network positions can be thought of as forms of inequality in social capital.
\cite{jackson2020} offers more background and references on social capital and how networks relate to various forms of it.  Here the focus is instead on how homophily in networks affects three types of interactions and how those drive inequality.}
The fact that homophily in networks ties younger cohorts of a community to older cohorts, leads inequality to reproduce itself across generations, with the same groups being disadvantaged over time.
That provides key insights behind the tight relationship between inequality and its persistence:  immobility.

This perspective on social networks provides a social capital complement to the usual financial and human capital inheritance stories.
Including both is vital since, for instance, in \cite{chettyetal2022I} we find that two thirds of
differences in upward economic mobility across communities
can be predicted by differences in a measure of social capital
that captures how little homophily there is by socio-economic status. The results are robust to controlling for standard
explanatory variables measuring education, income, ethnic composition, and household structure.
The structure of social networks plays a starring role in predicting mobility and inequality, and a very different role from financial capital and family.

The importance of distinguishing the variety of forces behind inequality becomes most evident when we turn to the discussion of policies
for overcoming inequality.  Some economic drivers can be addressed by redistribution, regulation, and subsidies.  However, those policies do not eliminate the social drivers and hence can leave
long-term inequality and economic immobility largely unaffected.  Resources spent in such ways can end up dissipating, with short-term alleviation of symptoms but little long-term impact.
Social factors and unequal distribution of social capital require policies that overcome informational frictions and divides in networks, and take advantage of the feedback effects and social multipliers inherent in networks.   These include mentorships, internships, affirmative action, and subsidizing education.  Lessons from network formation and homophily
also suggest important ways in which the structuring of schools and other institutions can be better designed to lower homophily and improve the reach of people's networks.

Key roles of social networks as a driver of inequality are discussed in an excellent antecedent to this paper by \cite{dimaggiog2012} (see also \cite{small2009unanticipated,dimaggiog2011} as well as
the afterward of the second edition of \cite{granovetter1995}). My focus is broader on some dimensions, including discussion of economic drivers of inequality, and more extensive on other dimensions, such including the discussion of the relationships between homophily, inequality, immobility, and productive inefficiencies.  I also discuss
the different policy implications of the variety of forces underlying inequality.

\section{Background}

Before proceeding to discuss the drivers of inequality,
it is useful to briefly discuss a few key facts about inequality, as well as why we care about it.

\subsection{Why Inequality Matters}

Discussing why we care about inequality helps focus our attention and to
frame a discussion of which policy interventions can be justified when.
The voluminous literature on the philosophy of inequality and welfare would take us on a tangent here,
but it is worth mentioning a few key points.

Given how resilient and widespread inequality is, it has been a central subject of
many studies in political philosophy, sociology, and economics over the decades.  The roots of modern thinking trace to prominent contributions include those by Jean-Jacques Rousseau \citeyearpar{rousseau1754discourse,rousseau1762social}, Freidrich Engels \citeyearpar{engels1845condition},
John Stuart Mill \citeyearpar{mill1859,mill1869subjection},\footnote{Ironically, although ``The Subjection of Women'' was published with John Stuart Mill as sole author, he acknowledged that some
of the text was inspired by writings and ideas of his stepdaughter Helen Taylor and that much of his writing on the subject  was influenced by extensive discussions with his wife Harriet Taylor Mill.
He also stated that his collaboration with his wife was extensive in the writing of ``On Liberty''.   }
Karl Marx \citeyearpar{marx1867kapital}, Henry George \citeyearpar{george1879}, Max Weber \citeyearpar{weber1921}, John Rawls \citeyearpar{rawls1971}, James Coleman \citeyearpar{coleman1974inequality},
Anthony Atkinson \citeyearpar{atkinson1975}, and  Amartya Sen \citeyearpar{sen1992}.\footnote{There are also a variety of influential works on specific forms
of oppression and inequality, such as key feminist writings (e.g.,
Mary Wollstonecraft \citeyearpar{wollstonecraft1792vindication}),  and a long string of prominent abolitionist writings including those by  James Ramsay, Harriet Beecher Stowe, and Frederick Douglass; not to mention
various discussions of religious and ethnic oppression throughout history.}

At its core, concerns about inequality are often based on some notion of fairness, as well as what we might refer to as social insurance.
People can end up in disadvantageous circumstances (for reasons beyond their volition\footnote{Unpacking the extent to which people are responsible
for their outcomes is an important subtheme in this literature, but it is clear that much of immobility is not self-imposed.}), and should not suffer from those disadvantages.
One of the most powerful tools in framing this perspective was articulated by \cite{rawls1971} and
is a hypothetical exercise where one imagines that one could have been born into any position of a human in the world.
As Rawls states, ``Among the essential features of this situation is
that no one knows his place in society, his class position or social status,
nor does any one know his fortune in the distribution of natural assets and
abilities, his intelligence, strength, and the like. ... The principles of justice are chosen behind a
veil of ignorance.''

One way to use the veil of ignorance to address inequality, is to ask what policies would we like to implement to
make the world the best possible if we look at the world from behind this veil of ignorance and realize that we could have ended up in any position in society.
However, while Rawls' hypothetical frames the discussion nicely in terms of endowments of natural characteristics, as well as the economic and social capital that one is born into,
it can still lead to a wide variety of views on what fairness means and how one should evaluate various policies.
One possibility is to view our position as a chance phenomenon, and say that we could equally likely have been anyone in the world (or even in future generations), and
then ask what policies maximize the overall average welfare of the society.     This is the
perspective of utilitarians (e.g., \cite{bentham1789,mill1859utilitarianism,harsanyi1977rule}), who weight the relative utilities or enjoyment of
the benefits of different outcomes across a population.
Interestingly, instead of looking at the outcomes for all positions as a utilitarian would, Rawls instead argued for concentrating on the welfare
of the worst positions.  I won't venture into that debate here,\footnote{See \cite{barberaj1988} for axioms that characterize Rawls' approach.}
but regardless of how one evaluates welfare from thinking behind the veil of ignorance,
it provides a justification for comparing the circumstances of different individuals and the consequences of those circumstances.

Most fundamentally, the veil of ignorance gives us a reason for a form of social insurance.  If the circumstances of one's birth substantially limit the opportunities and outcomes that an individual can expect,
it is too late for the individual to be able to insure against those circumstances.
There are many things that can lead to unequal outcomes.  For instance, some families have their house burn down and their possessions destroyed, while others do not.
This is something that people can privately insure against.
In contrast, being born into poverty and ending up with poor childhood health and education is not something that a person has a chance to buy private insurance against.
By most any definition of fairness operating behind the veil of ignorance, a society has a moral obligation to address those circumstances that
disadvantage an individual from birth.  The individual cannot insure against the event, but society
can.  It requires society to do two things:  one is to compensate people for the outcomes that are beyond their control, but the other is to ensure opportunities are more
evenly available to people regardless of the position into which they are born.\footnote{The details of what this entails are something that I will not delve into here,
but of course are very important. Allowing people to access a highly productive life rather than just one that passes a minimum standard of subsistence
can be a difference between changing opportunities and providing some ex post redistribution.  How does one account for things like free will and dignity?  Which outcomes and situations are ones that should be the concern of a society?  How much should it depend on choices and liberties afforded
to the individuals?  Various forms of diversity can be at odds with some notions of equality and liberty.  These are the subject of important further discussions, such as that of
\cite{sen1992}.}

Orthogonal to the moral underpinnings of fairness and social insurance, there are
also basic economic reasons for caring about missing opportunities and frictions in a society.
Missed opportunities lead to a loss of production for a society, which
leads to a smaller overall pie for the society.   Society loses when people's talents are under-utilized or mis-matched with their profession and other positions in society.
These losses can compound over time, leading to lower growth (\cite{perssont1994}\footnote{For additional discussion see, for instance, \cite{benabou1996inequality,stiglitz2012price}. The relationship between inequality and productivity
is challenging to identify  \citep{barro2000}, but recent data offers increasing evidence for the relationship and potential channels (e.g., \cite{berg2018,aiyar2020inequality}).}).
At a most basic level, inequality can lead to differences in health outcomes and stress (\cite{adler2002socioeconomic}).  Ill health for an individual, especially when it begins in childhood, compounds itself over time and is costly for society, both in the lost production and the resources used in the care for that individual.
Inequality can also manifest itself in things like differences in group participation and social capital (\cite{alesinaf2000,laferrara2002}), which also have productive consequences.
Inequality can also cause political frictions leading to less effective government, and also cause a society to focus
on redistribution instead of investment (e.g., \cite{voorheis2015unequal}).

\subsection{Inequality and Immobility}

As is clear from Henry George's quote above, systemic inequality has been endemic for centuries.
In fact, we see it going back millennia and across most societies---for instance, in the contrast between
the more than one hundred kilograms of gold that plated
King Tutankhamun's innermost coffin and the plain graves of the workers who built his tomb.  Lui Jin, a key courtesan during the Ming Dynasty, died with hundreds of thousands of kilograms of gold, while the much of his society's booming population had no property or savings at all.
Nicolas Fouquet's stunning Ch\^{a}teau de Vaux-le-Vicomte earned the envy and ire of Louis XIV and inspired the Ch\^{a}teau de Versailles, at the same time that repeated famines ravaged France.
John D. Rockefeller's ruthless monopoly amassed more than a billion dollars when the per capita income in the US was on the order of a hundred dollars a year.

As systematic data have become more available over recent decades, we have increasingly sharp and broad measures of the extent and dynamics of inequality and economic
immobility,\footnote{For instance, see \cite*{bourguignonm2002,bowles2002inheritance,atkinson2011top,piketty2014,chetty2017fading,corak2016,saez2016wealth,hubmer2020sources}.  Such measures
continue to evolve as reliable data for parts of the world are still difficult to obtain. but there are new methods of inference (e.g., \cite*{mirza2021global}).}
as well as their prevalence across societies and geography.\footnote{See, for example, \cite*{charles2005occupational,bowles2010emergence,corak2013,chetty2014land,atkinson2014chartbook,alfani2021}.}

Regardless of short and medium term trends and changes, inequality is a persistent phenomenon, and its spatial and demographic correlation patterns indicate that it is
not just a chance phenomenon  \citep*{chetty2014land,corak2016}.
From our perspective, a particularly important piece of the puzzle is that
inequality is closely related to economic immobility:  we can predict which people from the next generation will be most likely to end up at the bottom of
the income distribution.
This is particularly problematic when viewed from the perspective of Rawls' veil of ignorance.

\begin{figure}[h!]
\begin{center}
\includegraphics[height=4in]{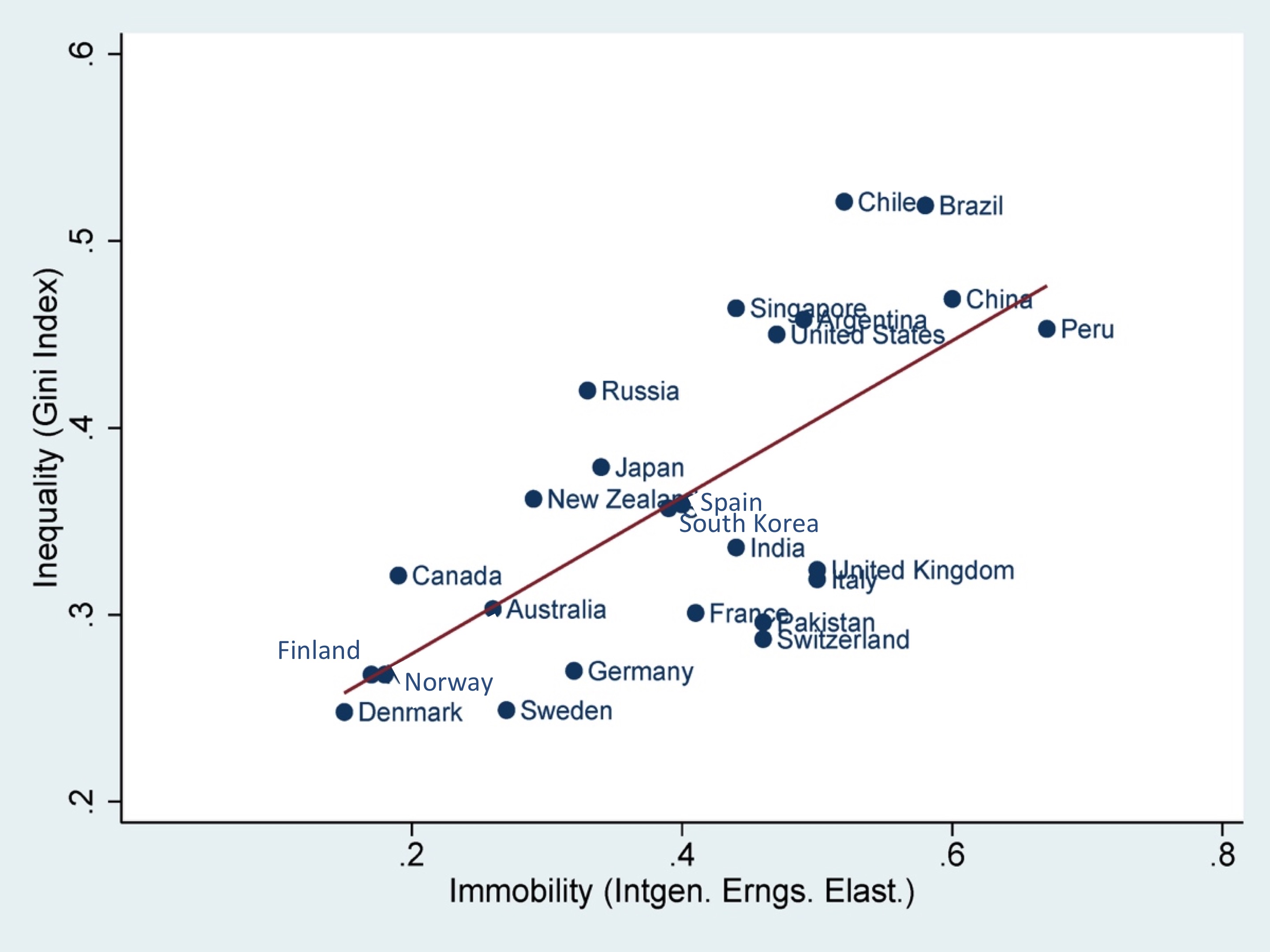}
\caption{\label{gatsby}  [The Great Gatsby Curve]
Inequality (Income Gini coefficient) plotted against economic immobility (intergenerational earnings elasticity).
}
\end{center}
\end{figure}

The relationship between inequality and economic immobility was crystalized by Alan Krueger in a speech in 2012 in which he depicted the ``Great Gatsby Curve,'' built on the work of Miles Corak.\footnote{See \cite{corak2013,corak2016} for more background, and \cite{andrewsl2009} for related observations.}
The observed fact is that countries with higher inequality also tend to have higher economic immobility.
An updated version of the Great Gatsby Curve appears in Figure \ref{gatsby},\footnote{Some data are from Corak \citeyearpar{corak2016}, supplemented with
data from the CIA World Factbook, \cite{borisov2016intergenerational}, and \cite{kim2017}.
I have reversed the axes from the original.  This reversal conceptualizes
immobility as the base force, and (recurrent) inequality as a result of it. }
and it depicts the relationship between inequality---as measured by an income Gini coefficient---versus economic immobility---as measured by
intergenerational earnings elasticity (the coefficient of a regression of log of child's earnings on log parents' earnings).

The correlation is observational, but growing evidence shows that there are causal aspects to it (e.g., see Section \ref{changing}).  Most importantly
from our perspective here, the concepts behind the Great Gatsby Curve help in guiding theory and policy.
People born into poor circumstances are disadvantaged in terms of their networks (social capital), as well as the financial and human capital of their families,
and this can lead them to end up with worse information and opportunities, which translate into worse outcomes.\footnote{For some measurements of the intergenerational persistence of human capital see \cite{adermondynastic2021}. }
The perpetuation appears as immobility, and the disparities across communities manifests itself as inequality, but they are both
features of the same phenomenon.\footnote{This relationship comes out of a variety of forces.  For instance, \cite{loury1981,benabou1993workings,benabou1996equity,durlauf2018understanding,chettyj2024}
discuss how it can come out of investments in education and neighborhoods across generations.  \cite*{bolteij2020} show how it comes from a model of homophily and  job referrals (more on this in Section \ref{social}).
These generally involve some social aspect:  family and community effects. }

\subsection{Homophily}

As homophily plays a starring role in the discussion of the social shaping of inequality, I give a brief description of the important aspects of it for our inequality discussion and
refer the reader to \cite*{mcphersonsc2001,jackson2019} for more detailed background and references.

Homophily is a relatively modern term,  coined by \cite{lazarsfeldm1954}, for the long-standing phenomenon of people associating with others who are similar to themselves.
To given an idea of its extent, consider an early empirical study by \cite{verbrugge1977}.  She found
substantial homophily on all the characteristics in her data, including:  age, years of schooling, profession, religion, marital status, and employment status.
In particular, she examined 240 different categories and examined odds ratios of whether
people named other people with the same characteristic as their closest friends.   As an example, she compared the frequency with which Catholics named a catholic as their
closest friend to the frequency that non-Catholics named Catholics as their closest friend.
A ratio of one corresponds to no homophily.
She found that the odds ratio for the category Catholic were 6.3 in Detroit and 6.8 in Alt Neustadt.
All of the 240 categories she examined had odds ratios above 1, and  225 of the 240 odds ratios were statistically significantly above 1---with odds ratios between 2.2 and 81.
Indeed, this study is just one example, and it is difficult to find dimensions on which human networks are not homophilous. Homophily even
extends to personality characteristics.\footnote{See \cite*{morellietal2017,jacksonnsy2022}.}
Obviously, many dimensions are correlated,
but the important fact for understanding the impact of homophily is that it leads to networks in which groups can be relatively isolated from each other socially, even when they
are in close proximity.

It is worth emphasizing that homophily exists for many reasons, and these become policy relevant as we discuss in Section \ref{policies}.\footnote{See \cite{jackson2019} for a more extensive overview and \cite*{currarinijp2009,currarinijp2010} and \cite{chettyetal2022I,chettyetal2022II} for some empirical background.}
In order to better understand the reasons for homophily and policies that can help mitigate it and its impact,
it is useful to decompose homophily into two parts.
One is that homophily is driven by exposure and contact patterns (\cite{allport1954}).
It is easier to form and maintain friendships with people who are physically or virtually close by.
For instance, students of the same age are grouped in schools together and sometimes even placed
in tracks based on their prior achievements,
people with similar education levels work together, and people of the same religion attend services together.  Housing prices often vary by the amenities and location of neighborhoods,
which can then segregate people by income.
Thus exposure is a primary driver of homophily, given people's heavy tendency to interact with those with whom they are around---and given that people
are often partitioned in physical and virtual spaces according to their characteristics and interests\footnote{There are many reasons for such separations
that I won't detail here, but see \cite{jackson2019} for more background and discussion.} ---this leads to substantial homophily.
The other part of homophily is people make choices about whom they interact, even conditional on exposure.
That is, in a room of people with diverse backgrounds, they will often gravitate towards those with shared backgrounds and interests.
People find it easier to interact with people with whom they share a common background, as they share history, expectations, and norms of behavior.
Sharing a background also correlates the challenges people face and what they can learn from each other.\footnote{See \cite{lobel2016preferences,aybasj2021} for  a models of the tradeoff
between shared preferences versus more diverse information sources.}
New parents of a certain income and education level share very similar challenges in raising their children.
Thus, homophily ends up being driven not only by the ways in which people are separated into communities within a society, but also people's tendencies
to choose similarity even when faced with diverse choices.\footnote{See \cite*{currarinijp2009,currarinijp2010} and \cite{chettyetal2022II} for
analyses that examine such decompositions empirically.  Both appear to be substantial, although it is not always easy to distinguish the two
from observational data, as some aspects of exposure patterns are not observed and are partly endogenous.  People make choices that change the sets of people with whom they interact: for instance if someone goes to a university that has people who are mostly similar to themselves, is that part of exposure or was that a deliberate choice as they wanted to have that exposure?}
This distinction between exposure and choices conditional upon exposure is important not only because informed policy decisions depend on what drove the networks to form the way they did; but also because
these different reasons have different consequences, as discussed in Section \ref{policies}.

\section{Economic Sources of Inequality}

Now I turn to discussing some of the sources of inequality, beginning with some of the most prominent economic ones.

\subsection{Unequal Returns to the Factors of Production}

A basic economic source of inequality is unequal returns to the factors of production.
This can come in a variety of forms.
As made popular by the studies of \cite{piketty2014}, for instance,
a greater return to capital compared to the growth of the rest of the economy,  ``$ r>g$,'' can lead those holding capital to have their
wealth grow more rapidly than the rest of the society.
The importance of capital and its potential role in inequality has roots in the writings of
\cite{marx1867kapital} and the literature that followed, but is examined from a different perspective in Piketty.

The approach of Piketty echoes
the theory that Henry George presented in his book  ``Progress and Poverty'' \citeyearpar{george1879},  which was one of the most widely sold books in the world up to the end of the 19th century.
It also concerns the idea that inequality can be driven by certain factors of production earning higher returns than others.
However, instead of capital, George's focus was on rent: the returns to owning land.  He noted that people owning land earned much more from its cultivation than the people who worked the land.
George works at explaining {\sl why} one should expect a disparity.
In particular, he asks (Chapter 11, \cite{george1879})  ``To say that wages remain low because rent advances is like saying that a steam-boat
moves because its wheels turn round. The further question is, what causes rent to
advance? What is the force or necessity that, as productive power increases, distributes a
greater and greater proportion of the produce as rent?''
The answer that George gives  stems from imperfect competition.
He states (Chapter 14) ``That as land is
necessary to the exertion of labour in the
production of wealth, to command the land
that is necessary to labour is to command
all the fruits of labour save enough to
enable labour to exist.''
and later
``Material progress cannot rid us of our dependence upon land; it can but add to the power
of producing wealth from land; and hence, when land is monopolized, it might go on to
infinity without increasing wages or improving the condition of those who have but their
labour''
A caricature of George's argument is that labor is supplied competitively, earning its subsistence value,
while land is monopolized and earns the entire remaining value of the fruits of production.

Piketty's core argument has a similar structure as that of George's \citeyearpar{george1879} core argument, substituting returns to capital for the rents earned from land.  Piketty \citeyearpar{piketty2014} also provides some other scenarios under which returns to capital could exceed returns to labor, as well as a detailed look at the accounting of inequality and a number of perspectives on its incidence, dynamics, and implications
for taxation.\footnote{
For a variety of comments and debate surrounding Piketty's argument, evidence, and conclusions see \cite{krugman2014,mankiw2015yes,jones2015,acemoglur2015,piketty2015}.}

Over time one can see a progression in which forms of capital are most vital and monopolizable.  As agriculture became a shrinking part of the economy
more of the returns moved to the financial capital and control of means of industrial production (e.g, the oil and railroad barons of the nineteenth centuries).
More recently, growth has moved increasingly to service industries, which rely less on financial capital  but
can still hold enormous economies of scope and scale.\footnote{There are also effects of globalization on inequality, as discussed, for instance by
\cite{krugman1995globalization}.}
For example, substantial growth in modern monopolies have come from social media, where the more users and content a given site has, the more attractive it becomes.
The profits to such monopolies have been enormous and very concentrated.  For instance, as \cite{atkeson2020understanding} show, one can explain much of
the large amount of wealth concentrated at the very top of the distribution, simply by accounting for families that make undiversified investments in businesses that happen
to become very large and profitable.\footnote{For more background on churn among the very wealthy, see
\cite{gomezdecomposing}.  The level of wealth that one has upon inheritance can also affect how wealth accumulates or is depleted across generations,
as found in \cite{nekoei2019inheritances}.}  Thus, some random chance, coupled with undiversified investments and extreme returns to some enterprises can also lead to a large concentration of
wealth a the top of the distribution, without any special returns to capital on average.
More generally, how the fruits of production are shared depend on property rights, regulation, and a variety of cultural norms (e.g., see \cite{mulder2009intergenerational,bowlessm2010}).

Nonetheless, as \cite{piketty2014} acknowledges (see also \cite{piketty2015}), much of the recent growth in inequality
is in income.   This can partly be tied to increased returns to education.
Thus, while excess returns to land, capital, and monopolies, can be important drivers of wealth inequality,
they leave much unexplained about the large amount of income inequality that still exists even when we completely ignore the wealthiest few percent of the population.

Let us next turn to some economic explanations for the sizeable inequality in employment and earnings that we see across the world and within many countries.  This inequality is substantial even when we ignore the highly wealthy; and includes the gap between outcomes for those with post-secondary education and those without.  It is a type of inequality that social factors have the most to say about.

\subsection{Frictions in Human Capital and Other Poverty Traps}

The ratio of the wages of those having a college degree compared to just a high school degree or less has more
than doubled in the last half-century.\footnote{See the background and references in \cite{jackson2019,autor2020}.  The wage gap is also mirrored by a growing longevity gap, as found by \cite{case2021life}.}
Again, a short caricature is that technological advances are making high-skilled labor more productive, while those same advances are replacing low-skilled labor.  That is,
technology in the form of improved computing power, communication, and automation,  have been complementary to high-skilled labor while substituting for low-skilled labor.

This is not just a modern phenomenon: it happened also during the industrial revolution when agricultural labor was increasingly displaced and returns to skills in manufacturing grew.
It happened again a century later when labor was displaced in manufacturing in many industrialized countries and shifted to services.
The natural adjustment of this should be that more people invest in education that makes them best-suited for the new productive opportunities.
In the current setting, this would drive up the supply of people with high levels of education and drive down the supply of relatively uneducated labor,
bringing the wages back into balance (adjusting for costs of education).\footnote{See \cite{tinbergen1975} for a classic discussion of this, and
\cite{goldin2009race} for more detailed discussion of the relative rates of technological change and education advancements.
See also the empirical analysis and theory of  \cite{kuznets1955economic}, who posited that growth and changes in production could lead to increased inequality as capitalists profit from expanding industries, while
competitive wages and some moving frictions lead to falling wages in areas left behind and slow wage growth in the new industries.  He also predicted an eventual decline in inequality once the period of change subsided.
Evidence for this Kuznets curve is mixed (e.g., see \cite{alfani2021}).}

There are many frictions here.
Some are geographic, as taking advantage of new opportunities can require uprooting which is both risky and costly---leading people with fewer resources to stay behind (e.g., see \cite*{connor2020changing}).
Some reflect the slow growth in availability of the supply of education needed.  For instance,
access to higher-education has been slow and in fact stagnant in many countries.\footnote{For example, see the statistics on investments in education and amount of higher education from the US National Center for Educational Statistics.  In contrast, since the 1990s higher education has been growing rapidly in China - although still not at a fast enough
rate to compensate for the rapidly increasing demand for qualified graduates.}
Some reflect the barriers to the necessary investment among poor families, and other related inheritance issues (e.g., see \cite{bowles2002inheritance,sacerdote2007large,benhabib2011distribution,fagereng2018wealthy,herbst2021opportunity}).
As education is cumulative, differences in parents resources', opportunity costs, and approaches to educating their children at early ages, as well as
available public resources, can be constraining (e.g., \cite{bowles1972schooling,lareau2011,falk2021socio}).\footnote{See also \cite{durlauf1996} who embeds such an education poverty trap within
a model in which families choose neighborhoods in which to reside.  Wealthier families choose to live in neighborhoods that are not
affordable to poorer families, for the greater financing of education as well as the peer dependence on educational attainment.}

This can be seen as an example of a more general form of a poverty trap (e.g., see \cite{carterb2006}), where
a lack of resources leads to under-investment and low returns, which leads to a lack of resources.  Effectively there
are increasing returns to investment, and without minimal investment, one cannot get out of poverty.  Such vicious cycles reflect
difficulties in borrowing, especially given the lack of collateral and the moral hazard and adverse selection issues associated with such lending;  not
to mention the early ages at which such investments are often most effective (e.g., see \cite{loury1981}).
There are related poverty traps in health, as poor nourishment and health leads to lower productivity and prolonged poverty, which
then lead to lower nourishment and health.
These sorts of poverty traps, and lack of adjustments to shifting technologies and economies, also have important social network/social capital aspects to them:
as adjusting takes both the knowledge of how to adjust, as well as the opportunities to do so, as discussed more in Section \ref{social}.

\subsection{Discrimination and Statistical Discrimination}

Another long-standing source of inequality, and one that appears in many forms, is discrimination.
An advantaged group favors its own members and discriminates against another group's members, such as according to ethnicity, social class, religion, gender, caste, language,  and other characteristics.
It has been wide-ranging across millennia and cultures, and has manifested itself in the forms of slavery, caste systems, clan systems, racism, segregation, sexism, stereotyping, and in the many
ways in which some groups are favored in access to education, jobs, and political positions.  There is extensive evidence of discrimination showing that not only is it widespread, but also that its
impact can be large.\footnote{E.g., see
\cite{newman1978discrimination,fix1993overview,fershtman2001discrimination,bertrand2017field,hangartner2021monitoring,huber2021discrimination}.}

Researchers have distinguished between two forms of discrimination: one is a direct culturally driven discrimination in its many sociological forms, and is sometimes referred to as ``taste-based''  \citep{becker1957economics,pager2008sociology}, and the other is often referred to as
``statistical discrimination''  \citep{arrow1971some,phelps1972statistical}.
A prominent theory for culturally driven discrimination is due to \cite{weber1978economy}: a group
socially closes itself in order to gain or maintain exclusive access to scarce resources.  It can use various identifiable characteristics to distinguish itself, and hence to discriminate against outsiders.
In contrast, under statistical discrimination, people distinguish others based on expectations that they will have lower human capital and productivity.
That expectation makes it more difficult for the discriminated group to get good employment.  Given the worse employment, members of that group then have less of an incentive to invest in becoming productive and so
the group ends up being less productive which confirms the expectation.
Statistical discrimination can be seen as another form of poverty trap, and a particularly pernicious one, since the group that is being discriminated against cannot
escape even if they are given the resources to invest in developing their human capital, as they are still stereotyped as being less productive.
These two forms of discrimination are not necessarily mutually exclusive, since group competition can lead a group to exclude out-group members, which leads them to invest less, which then makes the discrimination
statistically self-fulfilling.\footnote{For
more background and more discussion of the relationship and distinctions between these two forms of discrimination, and what evidence there is for each, see \cite*{arrow1998has,lang2012racial,bohren2019inaccurate}.}

\section{Social Drivers}
\label{social}

As mentioned above, inequality has been a central topic of the sociology literature
including major theories of social stratification and class distinctions, discrimination, and collectivism (groups having
priority over individuals), among others.
Rather than attempt to survey that enormous literature,
I take a narrow focus
that is more micro-oriented and operational, and examines the interplay between social networks and economic behaviors.
Many of these basic sociological forces end up driving inequality through the divisions in social networks that they induce,
and thus much of the policy discussion in Section \ref{policies} still directly, or indirectly, addresses those social forces.

I focus on three main ways that people's friends and acquaintances impact their behaviors and outcomes, and result in inequality and immobility.
I take each in turn.

\subsection{Unequal Job Opportunities via Social Connections}

The pervasive use of referrals and connections in obtaining jobs has been extensively documented.\footnote{E.g., see \cite*{myerss1951,rees1966,reess1970,granovetter1973,montgomery1991,granovetter1995,ioannidesd2004,topa2011,rubineauf2013,zeltzer2020}.}
The data from such studies show that connections are a primary channel through which people obtain jobs
across almost all professions and education levels, and across countries and cultures.

There are several reasons for which jobs are often filled via connections rather than open applications or other means.
One is that there is a lot of uncertainty about how an applicant will actually perform in a new job.
Are they sufficiently skilled?  Will they work well with others?  Will they be responsible and productive?  Will they stay for a long time?
Interviews and resum\'es provide very imperfect answers to these questions.  References and referrals provide much more insight into these questions (e.g., see \cite{bewley1999}).  Most importantly,
current employees and people well-connected to a company know what a particular job entails, and they are best-suited to find good matches for an open job.
For example, many applicants may have some background in coding a certain software and be literate enough to pass an interview, but the applicant
might still not have the skill to perform a project that the employer has in mind.  This is something that would be known by someone who is involved in the
project or knows it well, and has either worked with this person or overseen that person.  Thus, the personal connection becomes very useful.
Indeed, there is ample evidence that people hired via referrals are
more productive, creative, and stay longer in their positions than people hired via without referrals.\footnote{E.g., see \cite*{fernandezcm2000,brownst2012,fernandez2014causal,burks2015,dustmann2016,pallaiss2016,bond2019networks,benson2019discrimination}.}

A second reason for using referrals is that employers often wish to find people similar to the best of their current employees (e.g, see \cite*{fernandezcm2000}).
These employees have the right background for the position, and fit well with the time demands, and have not quit.
It can be hard to find similar people from looking at applications, whereas by using the social network
an employer can actually take advantage of homophily.  The demands of the job might not have anything to do with skill levels, but more basic
things like not stealing from the company, being willing to work long hours, or travel extensively, or work strange shifts, or be available for extra work at short notice, etc.
By hiring via the network they are more likely to
hire someone who is very similar to their current employees who have these characteristics.  This might not diversify their workforce, but often an employer's priority is finding someone who can do the job and will not quit too soon.

A third reason is that friendships with other workers can directly affect the performance of employees as well as their turnover---having friends within an organization can lead a person to perform better and stay longer (e.g.,
\cite*{fernandezcm2000,fiorillo2014job,brownst2012}).   On the flip side of this are forms of nepotism and favoritism, which can have negative effects \citep{ponzo2010use,bramoulle2016favoritism}, but those depend on the level of homophily \citep{horvath2014occupational}.

Regardless of the reason, when coupled with homophily, the reliance on referrals leads to feedback effects and can be a basic force behind inequality in both employment and wages.
If one group has higher employment than another group, in some industry or just in general, then due to homophily the first group's members can expect to receive more referrals than the other group's members, on average.
Getting more referrals leads to a higher chance of being employed, as well as being matched most productively.
The intuition behind this is straightforward: taking the best option out of a set gets better when that set is larger.
Having more options also improves a person's bargaining power leading  to higher expected wages and promotions.
For instance, as discussed by \cite{arrowb2004,okafor2020social}, relatively small differences in the number of referrals different groups
have access to can explain a substantial amount of the differences in wages by race.\footnote{People's positions in their networks affects their bargaining power and who they are neighbors with more generally, and can lead to quite unequal outcomes when there are strong asymmetries in position, as discussed by \cite*{ketsetal2011,chengx2021}.}
In addition, there is also a further feedback in that employment helps build more contacts, further enriching a person's network and opportunities (e.g.,  see \cite{granovetter1988sociological}).\footnote{Homophily can also affect the contacts that one has within a firm or other organization, which can affect one's chances for retention and promotion (e.g., \cite{opper2015homophily}).   It can also affect one's cultural fit within an organization (\cite*{stephens2014social}), which is known to affect hiring (\cite{rivera2012hiring}) but can also matter afterwards. }

Referrals via people's networks also end up correlating the employment and wages among friends, as analyzed by \cite{calvoj2004,calvoj2007}.  Having more employed friends
leads a person to have a greater chance of getting referrals, and multiple  referrals.\footnote{Employment over time can also affect the accumulation of friendships,
which can also have a further feedback effect (e.g., \cite{avinetal2015}).}
There is substantial (causal) evidence for this correlating effect of referrals in a variety of
settings.\footnote{E.g., see \cite*{sacerdote2001,munshi2003,beaman2012,patacchiniz2012,laschever2013,clausetal2015,beamankm2016,lalannes2016}. The wage effects are more subtle than
employment effects, and may depend
on details of the setting and the numbers of connections.  For instance, the analysis of \cite{arrowb2004} suggests that wages are concave in the number of connections,
which is seen empirically by \cite{berardils2019}.}

Moreover, since higher current employment of the members of some group leads to higher employment for that group's next generation
this also correlates employment and wages across generations, as for instance, analyzed by
\cite*{calvoj2009,buhai2023social,bolteij2020}.   Thus, the combination of homophily and heavy use of referrals in employment
can lead to immobility and provides an understanding for the strong relationship between
inequality and immobility.

In addition, the resulting inequality and immobility also have productivity implications.
The fact that people with poor connections have fewer chances to find jobs, and those with more connections can get multiple offers, means that the chances for people to be matched
to an employer is unevenly spread around the population (\cite*{bolteij2020}).
In general, if the value of an additional referral to a person in terms of matching
that person to the most productive employment is diminishing in the number of referrals that the person already has, then maximizing a society's productivity involves spreading the referrals
as evenly as possible throughout the population.
Instead, homophily leads referrals to be unevenly spread, and more heavily concentrated among
groups that are already well-employed.  Moreover, this can even further concentrate the referrals, as they end up
passed along to the relatively few unemployed among the advantaged group rather than to the relatively larger group of unemployed
among the disadvantaged group.   Thus, the combination of homophily and referrals  not only has inequality and immobility implications, but also average and total productivity implications.   Spreading
referrals more evenly throughout a population increases both the average level of productivity of those who are employed, and can increase the total amount of employment in the economy.

The combination of homophily and referrals can also distort investment incentives and result in a form of poverty trap.
For instance, a person who has, or expects to have, few employed friends has a low incentive to invest in acquiring skills to take advantage of employment opportunities.\footnote{E.g., see \cite*{calvoj2004,jackson2007,bowles2014group,bolteij2020,buhai2020social}.  It also affects network formation (e.g. \cite{calvo2004job,galeottim2014})  and career choices \citep{buhai2020social}, which then further correlate outcomes across groups.  In addition, people with fewer employed friends expect longer durations of unemployment over time \citep*{calvoj2004}, and so the duration of unemployment spells
can differ widely across groups \citep*{kailanr2021}.}
This is not only a question of being employed or not, but extends to many forms of investments in skills.  For instance,  someone who never expects to rise to a management position
has a low incentive to invest in acquiring the skills to become a manager.  This can also affect the expectations of employers, who then do not expect members of some groups to have
the backgrounds needed for some positions---which can then also play back into a form of statistical discrimination.

\subsection{Unequal Access to Valuable Information}

Job opportunities are just one form of information that flows through networks.  People also learn about many other things from friends, acquaintances, and various media.  This affects a broad set of
behaviors, including everything
from adoption of a fertilizer or new crop (e.g., \cite{conleyu2010,bandierar2006}), to participation in a microfinance program (e.g., \cite*{banerjeecdj2013}), a vaccination program (e.g., \cite*{banerjeecdj2019}),  a tax-deferred retirement program (e.g., \cite{duflos2003}),  to teen pregnancy \cite{kearney2015media} and education decisions (e.g., \cite*{sacerdote2011}).
Having networks that provide good information can be extremely important in determining people's lifetime health and well-being trajectories and outcomes (e.g., \cite{loury1977,bourdieu1986,loury2009}).

One might imagine that technological advances might overcome unequal access to various forms of information.  However, the information that people have and the opinions they form are heavily dependent upon their social networks, even in a social media-rich world.
A powerful example of this comes from \cite{bakshy2015exposure} which shows that exposure of people to cross-cutting political content is lowered not only by the
homophily in their networks, but then also by what is fed to them via a platform, and additionally by what they pay attention to.  When put together, in their data
only about a quarter of the content that people end up viewing is from sources with different political views.
That study examines news content, which can be shared broadly.  Information concerning investing in education, pursuing a career, and other large life-choices, is often
complex and nuanced, and learned from friends over long periods of time.  Given the strong levels of homophily among friends, this can lead
to even more inequality in access to information about such important investments.

The role of homophily is nuanced.  Homophily is a double-edged sword in social learning settings \citep*{lobel2016preferences,aybasj2021}.  For instance, as discussed above, a person can learn more from people who are similar to themselves.  A student who is from a low income family and has limited
preparation for university learns more about what attending university requires from others who have similar backgrounds, than from someone from a wealthier background in a special preparatory school.
On the other hand, given that far more people from wealthier backgrounds attend university \citep*{chetty2020income}, homophily can also disadvantage poorer people in terms of the number of people whom they know with
any university experience who show them whether attending college would be beneficial to them.
Ultimately one needs not only access to high quality information from closely matching peers, but across a variety of options.
This is the foundation behind some mentorship programs, in which people are connected to others outside of their own friendship circle
and who can provide valuable advice, but also whom the mentees can relate to, understand, and emulate (e.g., see \cite{putnam2016our}).

Importantly, homophily means that information about key opportunities and decisions can be distributed unequally across a population.
Without awareness of the availability of various choices and the costs and benefits of those choices, and enough information to be confident in a decision, people miss out on valuable opportunities.
For instance, if people of one ethnic or social group participate in higher education at a greater rate, then friends and families of those people gain more information about that experience and its benefits.
This then feeds back to lead next generations of the group with more experience to invest at greater rates, and the other group to continue to under-invest, so that there
is a feedback across cohorts \citep{aybasj2021}.\footnote{It is worth emphasizing that informational disadvantages can be particularly problematic for the very poor.  For them, uncertain outcomes---even with high expected values---can be too risky to
take advantage of since the costs of failure can be relatively enormous (e.g., \cite{banerjeed2011}).}
Again, this suggests that divisions in the network not only lead to inequality within a generation, but transmission of that across generations.

This is born out in strong empirical evidence that networks of connections across economic boundaries predict economic mobility.
For instance, we find in \cite{chettyetal2022I} that the economic mobility of low socio-economic status  people in a
given community is strongly predicted by the extent to which low socio-economic status people are connected to high socio-economic status people.
This is what the theory would predict, given that information and opportunities are more often in the hands of the high SES people, and so greater access to those people provides
increased chances of having better economic outcomes.
In fact,
the extent of these cross-income connections within a community (strongly) out-predicts  other standard community measures of poverty, segregation, social capital, and human capital, when it comes
to explaining economic mobility,\footnote{For additional evidence on how the geography and long range of people's networks impacts many behaviors see \cite*{bailey2018social,jahani2022origins}.} and in fact remains a strong predictor
when one includes other usual variables on a community's poverty level, ethnic composition, and education levels, etc.\footnote{It mediates those relationships.}
Interestingly, we also find evidence that these cross-income connections are at the heart of the Great Gatsby Curve.  In particular, once we control for a community's cross-income connections,
the relationship between inequality and immobility disappears (see \cite{chettyetal2022I} Table 2 regressions on the Gini coefficient).

Different types of social capital can have different effects on learning and behavior.  Having a tightly-knit community on a local level
can help that community function in providing incentives for people to cooperate and share with each other.\footnote{See, for instance, \cite{coleman1988} and \cite{jacksonrt2012}. See also \cite{ambrus2014consumption} for the importance of expansion properties of a network.}
However, that local support and clustering can be insufficient to help in economic mobility, as a very cooperative and supportive group can still be missing vital information
and opportunities that it needs to succeed.\footnote{In fact, there are reasons for which such tight-knit structures impede mobility,
as people may have to leave their groups and their support in order to take advantages of opportunities (e.g., \cite{austen-smithf2005,munshi2009mobility,barron2020too}).  However, on the other hand, having tightly-knit groups can be effective in improving
local public goods provision and collective action (e.g., see \cite{ostrom1990}). }
Gaining that information and those opportunities can require cross-cutting relationships, which is a very different form of social capital, and
social network feature, which is what we see in the data in \cite{chettyetal2022I}.
In fact, their support within the community turns out to be more advantageous in settings in which there is integration across socio-economic lines, so the two appear to have some complementarity.
More generally, the connectedness of people to information also depends on the quality of the local institutions that they are a part of,
such as child-care centers, and the outside resources and connections that such institutions offer and which vary widely across geography (e.g., see \cite{small2006neighborhood}).
Thus, one can think of economic connectedness as a feature not only directly of individual networks, but also of the networks of institutions through which people interact.

It is also worthwhile distinguishing two types of consequences of homophily and learning in networks.
One we can call passive.  This refers to the fact that homophily can lead different groups hold different beliefs over time, and to be slow or never converge to a consensus \citep{golubj2012,lobels2016}.
The other we can call active.   This refers to the feedback effects in actual decisions, as for instance outlined by \cite{aybasj2021}.
The fact that a group does not take advantage of some opportunity leads that group to learn less about that action and then be
less likely to take that decision in the future.   The active part involves feedback and a lack of exploration,  while the passive part refers to the lack of learning due to having less information to exploit.  Both matter, but can have different policy implications, as overcoming active learning consequences (e.g., a
group's lack of role models who can illustrate the benefits of education) can require policies like affirmative action or mentorships, while passive issues (e.g., a lack of information about the benefits and risks of some vaccine) can require supplementing information and contacts.
Given that many decisions and opportunities carry economic benefits, those benefits can be unevenly distributed across the population due to a lack of learning, leading to inequality and immobility patterns that
match the homophily.
And, in parallel to the job contact situation, many behaviors can be suboptimal.  People miss out on valuable decisions---such as educating themselves---that have consequences not only for themselves but also for
the overall productivity of the society.

Note that many different forces can all operate at once.
Having a productive life requires the proper training and development, the information of how to put that to its best use, and the opportunities to use it.
Economic and social hurdles hinder education and development,  homophily limits the extent to which information reaches a group and to which that
group tries certain options, and then homophily and the roles of networks in referrals and advancement hinder employment and advancement even conditional upon having the right skills and match
for a job.  All of these complement each other, and that complementarity has implications for how combinations of policies can be more effective than policies in isolation, as discussed more below.

\subsection{Norms, Culture, and Peer Influence}

The clustering of behaviors can occur for reasons beyond information sharing.
For instance, children imitate siblings and peers, people adopt fashions and fads, people
prefer to see movies and read books that they can talk to their peers about.   The many pressures
that push people to similar behaviors, can shape very consequential decisions
including pursuing education (e.g., see \cite{austen-smithf2005}).

Once again, homophily plays a starring role, as splits in the network can lead to different norms of behavior across homophily divides
(e.g., see \cite{jackson2007,golubj2012,jacksons2018}).
As a simplest example, imagine that people choose to dropout of school if a majority of their friends do, but not if a majority of their friends stay in school.  So, they simply prefer to do the same thing as the majority of their
friends - the simplest version of a game on a network (see \cite{jacksonz2014} for background).
In many networks it is easy to find (many) splits of the network into two groups such that each member of a group has a majority of their friends within that same group.
Then it is an equilibrium for everyone in one group to stay in school and everyone in the other group to drop out.

This is a stylized example, but the phenomenon is clear:  homophily leads to strong divisions across a network so that different groups have most of their connections within that same group.
This can lead groups to have different norms of behavior and cultures, even when they might be in close proximity to each other, and even have social connections with each other.
For instance, the friendships in many schools splits along ethnic lines, and that can very different behaviors across groups even within the same school.
These effects can be amplified by the fact that people can misperceive behaviors in the own social circles to be more representative of norms overall than they really are
(\cite*{jackson2019,jackson2019b,bursztyn2020misperceived,frick2020misinterpreting}).

It is also important to remark that peer effects are not only passive, but sometimes can be active.
That is, people may deliberate push their friends to act in certain ways:  students may bully students who do not conform to a group norm \citep{austen-smithf2005}, or people may pressure others until they
behave in certain ways (e.g., see \cite{calvoj2010}), or try to lead by example in hopes of getting others to act (e.g., see \cite{acemogluj2014,jimenez2021social}).
This can lead behaviors to become even more governed by network divides.

Note that many behaviors that are influenced by networks are cumulative.  Education, health, and aspirations can have cumulative
effects, so that as deficiencies become entrenched over time it becomes harder to eliminate them.  Moreover, the peer effects and lack of information that can lead to poor study habits, health habits, and low aspirations, feed back through the network both within and across generations to amplify these deficiencies.

\section{Policies to Overcome Inequality}\label{policies}

The variety of factors that can lead to inequality in outcomes suggest a corresponding variety of policies to deal with that inequality.\footnote{See \cite{page2018model} for a discussion
of using many models to both understand effect sizes as well as the limitations of focusing on just one framing of an issue, especially inequality.}
The policies to overcome the wealth concentrated in the hands of the top percentiles
of the population are different from those needed to deal with widespread and persistent differences
in health, education, employment, and income by ethnicity, caste, and gender that exist in many countries.
Nonetheless, one does not need a different policy for each issue.  Although more than one feature of social structure matters and more than one outcome is affected, understanding common sources and patterns can help us focus in on a few policy combinations, and can help us see that a common
network driver---homophily---is responsible for many differences in outcomes across groups.
Let us consider the logic behind various policies.

\subsection{Redistribution:  Addressing Symptoms or Causes?}

A class of policies often associated with combatting inequality is redistribution.
Such policies comes in many forms: progressive taxes, subsidies, universal basic income, price and wage controls, as well as government-run retirement, health, and welfare programs.
Effectively, these are all programs that re-allocate resources across the population.
Such programs reduce inequality to the extent that net benefits are seen by poorer individuals and net costs are seen by wealthier or higher income individuals.

It is important to distinguish two different reasons for redistribution.

The first comes from Rawls' veil of ignorance, which justifies redistribution as a form of social insurance.  By reallocating
resources from those in better circumstances to those in worse circumstances, a society is essentially
offering a form of insurance when viewed from behind the veil of ignorance.   The question of how much redistribution there should be, and who should benefit and who should pay the costs, is something that elicits a
variety of views.  What is viewed as optimal traces back to the moral foundations of society's responsibilities to its individuals versus their responsibilities to themselves.  This is well illustrated
in Sen's \citeyearpar{sen1992} discussion of the ``equality of what?''\footnote{Beyond the moral
issues, there are also a variety of incentive issues as such programs can distort people's behaviors.  For instance, for an illuminating discussion of the incentive issues
associated with equal sharing of resources, see \cite{abramitzky2011,abramitzky2018}.}

The second justification for redistribution is to combat poverty traps.  For example, if people cannot afford education, and cannot borrow privately to pay for it,\footnote{There are
many impediments to borrowing based on uncertain future values in the absence of having collateral to post, including moral hazard and adverse selection issues,
as well as the fixed costs of administering relatively small or highly idiosyncratic loans. }
then having a government provide it and/or help pay
for it can overcome the market imperfection.
The same is true of other forms of government sponsored loans and programs that spur investments in human capital and health.

This is an important distinction, since social insurance is essentially an ex post remedy to deal with the symptoms of inequality.
On the other hand, a justification for redistribution based on overcoming market imperfections in which people are trapped in poverty is aimed at overcoming causes.
Policies aimed at the causes
also deal more directly with immobility and inefficiencies in production.

Both justifications are important.   In treating a disease, one treats symptoms as well as causes---treating inequality is similar.  Nonetheless, the distinction is vital, since much
attention is focused on remedies for the symptoms and some of those will
not eliminate the causes of inequality or its longer-run effects like immobility, but instead will only overcome some of its immediate negative impact.
Some taxes, especially those aimed at wealth and inheritance, fall into both categories, as do regulations that combat monopolies and other forms of market driven inequality,
as they not only redistribute but also undo some of the sources of inequality (e.g., for a discussion of some of the impacts, see \cite*{chetty2009sufficient,dworczak2019redistribution}).

Policies can also be distinguished in terms of how leveraged they are.
For instance, it is investments in early education---even at preschool levels---have generally been found to large snowball effects.  Especially, early gains in non-cognitive skills help children gain further knowledge and skills.\footnote{E.g., see
\cite*{currie2001early,garcesct2002,heckmanetal2010,huggett2011sources,heckman2012,aizerc2012,heckman2014economics,felfe2018does,fryer2020introducing,kulic2019social,bailey2020prep,carneiro2021intergenerational,garcia2021lasting}.
Note however, that the research is not uniform in its findings (e.g., see \cite{durkin2022effects}).}  Some such policies more than pay for themselves when one estimates the social benefits compared to the cost of the policy (e.g., see \cite{hendren2020unified}).
It is worth noting that many aspects of children's education are determined by parents (e.g., see \cite{lareau2002,becker2009,lareau2011,doepke2019economics} ),
and thus educating parents about how to parent that can be an important part of an effective policy.  Evidence for this point dates at least to some of the findings from the original Coleman Report (``Equality of Educational Opportunity,'' 1966) regarding the importance of various peer
effects, local culture, and community factors, and has been studied in detail (e.g., see \cite{small2010reconsidering,gelber2013children,almond2018childhood,agostinelli2020takes}).
Moreover, peer effects among parents  mean that there network effects that should be addressed and leveraged by such policies.

\subsection{Addressing the Social Causes of Inequality: Leveraging Network Effects and Overcoming Homophily}

Understanding how homophily affects opportunities, information, and behaviors, provides
specific insights into constructing policies to overcome the resulting inequality, immobility, and production inefficiencies.

There are two different approaches that can each be effective in its own way.
One is to change the networks to try to lessen homophily and eliminate some of its perverse effects.
The other is to overcome the effects of homophily by providing opportunities, information, and changing damaging norms of behavior.
Any form of policy can have unintended consequences, and social engineering can be especially dangerous, so it is important to
consider how each of these works, and how the costs and benefits of each type of program work.

\subsubsection{Network Effects and Social Multipliers}

One aspect of networks that it is especially useful in policy design is that they are full of externalities and involve feedback and cascading effects.
On the negative side, these are at the heart of the causes of inequality and immobility, but these effects also work in reverse
can make well-designed policies much more effective.

To illustrate the point, consider a policy of affirmative action in overcoming unequal access to job opportunities.
As discussed by \cite*{bolteij2020} (see also \cite{granovetter1995}), the placement of people into jobs not only benefits those people directly, but also helps their networks of friends (and their friends, ...).
They can then offer referrals to others, as well as information that can help their communities better prepare for particular careers.
Thus, such a program should not be evaluated just on its direct effects on the individuals involved, but what it means for others within their communities.\footnote{Those indirect effects can be muted if, for instance, people placed into positions via affirmative action do not provide referrals and information to others who are similarly disadvantaged (e.g., see the discussion in \cite*{bolteij2020}).}
Role model effects are often mentioned in such a context, but there are more basic network effects that the individuals involved can provide access and information to their
friends, family, and acquaintances, well beyond simply showing that it can be done.   Moreover, because networks of relationships expand outwards exponentially, these effects can be substantial, even if the direct bilateral peer effects are small.  It also means that a temporary program can
have long-lasting impact:
giving one person a job that they would not have otherwise had can lead them to help others,
who can then help others, etc.\footnote{Such social multiplier effects are not just present in job markets, but in many other contexts too,
such as  crime  \cite{glaeserss2003}, tax evasion (e.g., \cite{galbiati2012tax} ), and smoking (e.g., \cite{cutler2010social}), to name a few.}
$^,$\footnote{There are also many other issues associated with such programs in terms of who bears the cost of short term implementation and
potential loss in productivity, how they affect people's perceptions of different groups, and the difficulties in changing programs over time (e.g., \cite{jehiel2021affirmative}).}

This perspective implies two things.  First it means that one should evaluate network-designed programs based on their full impact.  Second, it means that one
should take advantage of network structure in optimally targeting a policy.

To see this second point, consider a network policy in the presence of peer effects.
The causal evidence for such peer effects is now quite extensive over many domains  including, for instance,  exercise (e.g., \cite{araln2017} ), vaccination decisions (e.g., \cite{banerjeeEtalVaccine2021}), voting (e.g., \cite{dellavigna2016voting}),
and  education (e.g., \cite{sacerdote2001,hasan2013mechanics}).
To understand how
network structure matters in the presence of such peer effects, consider a specific application of students in high school and their decisions ow whether to continue to post-secondary education.
They can pay attention to their peers for many reasons.  One is that they simply enjoy behaving similarly, as that provides more commonality in experience and makes them feel as if they belong,
and it can also be that they learn from their peers, or that they see more benefits from continuing education as more friends do (e.g., from future job market contacts, etc.).

To make things concrete, suppose that students prefer to choose the same behavior as the majority of their friends.
Consider some community in which students all choose to stop their education at some point---essentially, stuck in a bad equilibrium.
If we have a limited number of scholarships (or other types of help) to give out to encourage students to continue their educations, it makes a big difference as to how we place these within the network.
If the scholarships are randomly scattered around the network, it can be that we only get the direct effect.   Those students go on to secondary education, but none of the other students have a majority of their friends continuing their education.
If instead, we carefully seed the scholarships within cliques and subgroups - so that they are concentrated and near each other, then they can lead other students to have a majority of their friends continuing.  This then leads those
friends to continue, and this can cascade outwards.\footnote{See \cite*{calvoj2004,jackson2007,jackson2019,gall2019college} for more examples and discussion of this point.}
Indeed, as \cite{jacksons2018} show, the benefits to careful targeting such seeds in the presence of peer effects can lead to arbitrarily large advantages compared to random placement.\footnote{For additional nuances see \cite{jacksonz2014,efferson2020promise,ye2020,galeotti2020targeting}.}

This is reflective of differences between the diffusion of basic awareness compared to norms and complex behaviors (e.g., see \cite{golubj2012,centola2018,jacksons2018}).
To be aware of a program subsidizing education might only require hearing from one friend, but to know how to take advantage of that program and actually participate in it might take observing several friends
go through it.
Thus, in terms of how one injects information and opportunities into a network to overcome inequality is not only governed by the shape of the network,
but also by how different sorts of behaviors and outcomes are shaped by the network.

\subsubsection{Changing Networks and the Challenges of Social Engineering}
\label{changing}

Let us now return to discuss the two approaches to overcoming the effects of homophily,
beginning with shaping networks.
Even though homophily operates differently through the opportunities, information, and norms that communities have,
homophily is a common force that correlates barriers to advancement on many fronts at the same time.  This means
that there is an advantage to addressing homophily directly.

The impact of changing a person's network can be substantial.
Some of the most eye-opening evidence of this comes from the impact of the ``Moving to Opportunity'' program, in which some families were subsidized
to move from poor neighborhoods into wealthier ones.  As shown by \cite*{chettyhk2016} the long-term effects can be enormous, especially when children
are moved at an early age. There were sizeable effects on health, education, and long term income, and they estimate, for instance, that
an eight-year-old child who moved saw an average lifetime earnings increase \$302000.\footnote{For another important example of this sort of effect,
see the analysis in \cite*{abramitzky2019intergenerational} regarding the importance of location in the mobility of immigrants.}
An interesting aspect of this effect is that it decreased with the age of the child.  Older children are more embedded in networks and behaviors, and both are
both harder to change when they have accumulated to a greater extent.

The formation of networks of friendships and other relationships are heavily driven by whom a person is in contact with (\cite{blau1977macrosociological,blaus1984,blau1987}).
If people have substantial contact with others who have access to opportunities and information, then they can take advantage of those connections.
By moving a child out of a poor neighborhood and into a wealthy one, all of the things we have been talking about are changed:  information flows, opportunities, and norms of expected behaviors.

As \cite{chettyj2024} show, homophily can also lead families to segregate by income and educational background, which then leads to differences in children's access to social capital.  This affects their networks and has the many consequences discussed above.   Such separating forces can be quite powerful and help explain why many families do not move to neighborhoods with improved educational outcomes for their children even when given the financial resources to make such a move, and even if they are altruistic and fully aware of the children's benefits.

Although experimental programs like Moving to Opportunity show what is possible,
it can also be that
shifting large groups of people around is not practical, at least in the short run.
Moreover, it might not even have the desired effect.  Moving a few families into different neighborhoods and seeing them integrate is very different
from making large-scale changes.
In particular, just putting people in contact does not necessarily alter their networks, unless they are a small minority who have little choice but to integrate.
One group having contact with another is necessary for forming cross-group connections, but can be far from sufficient.
For example, as found by \cite*{currarinijp2009,currarinijp2010},\cite{chettyetal2022II,mosleh2021shared} both exposure and choices conditional upon exposure matter to a sizeable degree in friendship formation.
In fact, some of the largest and most racially-balanced high schools in the US are some of the most fragmented along ethnic lines,
while smaller high schools are more integrated (\cite*{currarinijp2009,currarinijp2010}).

An example of how a policy designed to help individuals by engineering their communities can lead to unintended consequences comes
from \cite*{carrellsw2013}.   Entering cadet classes in the US Air Force Academy
were carefully structured to match together students who had the lowest predicted grade point averages with students who were expected to perform well.
Prior data had shown that cadets with low-predicted achievement performed above predicted levels when they had greater exposure to higher-ranked peers.
What was unexpected is that when the experiment built whole classes of roughly half very low-predicted achievement and half high-scoring cadets, then the cadets would self segregate.
Just as discussed above, when groups have larger critical masses, the friendships are more likely to splinter.  So although exposure of low-predicted achievers to higher-predicted achievers was optimized, the choices conditional upon exposure led to high homophily.
The idea behind this effect is intuitive.  In small communities, people have relatively little choice over whom they are friends with---groups are forced together.
When communities become larger, each group reaches a critical mass so that people can sustain a natural number of friendships just within their own community.  People's proclivity towards homophily
can take over and people have plenty of other people who are very similar to themselves with whom to be friends.
What happened in the Air Force Academy experiment was that instead of having small numbers of low-predicted achievement cadets integrating with high-scoring cadets,
they provided the low-predicted achievement students with many more opportunities to form friendships among themselves, leading to more segregation and worse outcomes.\footnote{It could also be that the high-scoring cadets having a critical mass allowed them to form friendships that excluded the
lows, as discussed by \cite{golub2021games}.}

Thus, although reshaping networks can be incredibly powerful at lifting people out of poverty,
these sorts of nuances mean that social engineering is tricky and can have substantial unintended consequences.\footnote{People can be involved in many networks at the same time, and changing one can end up affecting others, as can changing things that interact with them
(e.g., see \cite{banerjeeetal2020,atkissonetal2020}). }
Nonetheless, research to date suggests several insights that can be helpful in shaping successful policies.

One is this fact, just mentioned above, that smaller communities with a given composition of
groups foster greater integration across groups than larger communities with that same composition.
It suggests a new approach to lowering homophily can be to
improve the structuring of large organizations and communities by reshaping them into relatively small and well-mixed sub-communities.  Many aspects of urban planning as well as the structuring of universities, companies, and other
large institutions,
are aimed at trying to create smaller communities within them.  This suggests that not only can this be done to create tight community structures within larger organizations, but
also to help build bridges across different demographic and socio-economic groupings.
If a high school has huge numbers of students, then putting them into smaller communities who
take many of their basic classes together, and take meals together, etc., is like putting them in a smaller school where we know that more
cross-group connections tend to be built.\footnote{This also can tell us why some of the initiatives in the U.S. in the late 1950s through the end of the last century to build large schools that were well-integrated, via busing and redistricting, failed to lead
to well integrated friendships and interactions, and the desired improvements in achievement.  Without structuring the internal workings of the schools to foster tight small-group interactions,
students could self-segregate within their new schools just as they were segregated across their old ones.  The Air Force Academy experiment is a warning, however, that creating small groups that are too bimodal can still lead to segregation, even in relatively small cohorts.  }

Second, the network effects that we mentioned above mean that providing
new opportunities to diversify connections across a network has a sizeable impact.  In line with this prediction, our analysis in \cite{chettyetal2022I} finds
a strong relationship between economic mobility and increasing connectedness.\footnote{To put the slope in perspective,
changing
from the 10th to 90th percentile in economic connectedness predicts a 10 percentile increase in the income rank at age 35 of children from low-income families
(which is comparable to the estimated gap in upward mobility between Black and White Americans).}
The roughly linear relationship that appears in the data may reflect that each marginal bit of connection conveys similar benefits, or it may be that there are underlying
threshold effects that are smoothed out when averaged across a community.  Further exploration of economic connectedness may help uncover which aspects of the connectedness--- information, opportunities, or peer effects---are empirically most relevant.  Regardless, it is clear that cross-income connections hold promise as a key to improving economic mobility.

Third, simply increasing exposure can also affect discrimination and racism \citep{allport1954}, and recent studies show that contact can causally impact people's views and longer-term prejudice and behaviors \citep{rao2019familiarity,strother2021college,bursztyn2021immigrant}.  This can be true independently of the resulting friendship networks.  Moreover, contact can result in nonsocial and weak ties that are useful conduits of
information and opportunities even if they do not become strong relationships (e.g, see \cite{small2009unanticipated}).

Fourth, there are benefits from homophily that can be undone if people's networks are disrupted, and so it can be better to supplement rather than to dramatically rewire existing relationships.  Tearing communities
apart, without ensuring that new ones provide all the support that people need, can have damaging consequences (e.g., \cite*{barnhardtfp2016}).

Fifth, as technology becomes increasingly instrumental in shaping people's networks and their information, the design of
the algorithms that steer networks, exposure and information is becoming increasingly important.  Competition between platforms and profit motivations can
lead platforms to steer people towards connections and content that most closely mirror their backgrounds.   Even though such algorithms can enlarge the size of people's networks, it can also end up
increasing homophily rather than decreasing it \citep{jackson2019}.  Seemingly small design choices behind the algorithms that people use for communication, connection, and search, can have profound consequences.  Moreover,
the incentives of the designers are not necessarily aligned with greater social good, especially given the large network externalities at work.  Better understanding the workings and impact
of these services, as well as the incentives of the designers, is an important input into future policy design.

\subsubsection{Overcoming Network Structure}

Next, let us consider overcoming the effects of homophily and network structure without altering the networks, but instead
by counteracting their effects.

Some of the policies that help overcome the effects of networks are naturally suggested by the limitations that homophily imposes.
In particular, to overcome the lack of information and opportunities that people have due to not being well-connected, one can directly provide
the missing information and opportunities.   This provides a logic for mentorships, role models, and affirmative action policies.\footnote{All of these policies can also be thought of as reshaping networks to some extent, but that is often less their direct intention.}
As mentioned above, the network effects lead the information and opportunities to diffuse more widely and
thus can result in far-reaching and long-lasting effects (\cite*{holzern2000,miller2017,bolteij2020}).

The informational barriers in job markets can also be overcome by helping organizations find good matches without having to rely on referrals.
This can come in various forms.  One is subsidizing internships for relatively disadvantaged groups, so that firms get to know the workers.
Another is to provide better information about people's skills, by programs that allow them to show that they have reached certain skill levels.
In this direction, there is some hope that the proliferation of online education and certification will help close some gaps.
A less obvious factor in overcoming informational barriers is labor market rigidity.  If it is very hard to fire a worker, then the initial
hiring decision becomes even more important and referrals become even more valuable.  This can lead to increased inequality.  Thus, regulations that
are intended to provide job security can actually have a side effect of increasing inequality (e.g., see the analysis in \cite*{bolteij2020}).

Some policies, such as mentorships and affirmative action, can end up counteracting all three of the negative effects of homophily that we
discussed above.
Not only can these policies end up helping information and opportunities diffuse through the network, but
also, as people from given group end up having successful careers, they also serve a prominent examples
for their communities and can help reshape the norms, expectations, and culture within the community (e.g., \cite{acemogluj2014}).
These effects can also have increasing returns, at least initially.  That is, each additional success can
make the possibilities more visible and change people's expectations and aspirations even more, which
also have further peer effects.\footnote{Again, this is a place where age can make a difference, and earlier
interventions can be even more effective, and role models can have a wide impact on education attainment and trajectories (e.g., \cite*{beaman2012female}).}

\subsection{Policy Cocktails}

As argued by \cite{alfani2021}, ``human agency''---with an interpretation of the implementation of deliberate policies to overcome inequality---is a main predictor
of the trends of inequality over time; and when economies are left to operate unbridled, resources tend to accumulate
highly unequally.\footnote{See also, for instance, \cite*{hubmer2020sources} who find that changes in tax progressivity explain rises in wealth inequality.}

As we have discussed here, there are many forces behind such inequality and immobility, both economic and social, which suggest that a coordinated policy cocktail will be much more effective
than any policy in isolation.
Moreover, some policies deal with symptoms while others deal with causes, of which there are many,
and both causes and symptoms need to be treated.

A policy cocktail to fully address inequality requires at least four ingredients.
First up are the basic safety net policies and redistribution that provide social insurance and address the symptoms, or in an analogy to the treatment of a disease, overcome the pain.
Second are the long-standing economic policies that address the large wealth imbalances that accumulate from monopoly rents and increasing returns to various
forms of capital.  These include targeted regulation and taxation to address these issues without damaging economic incentives.\footnote{There is large literature on the structuring of various programs,
and how that can affect incentives as well as how easy it is for people to access a program (e.g.,  \cite{mirrlees1971exploration,katz1990impact,krueger2002labor,currie2006take,chetty2008moral}).
I don't address those issues here, but instead focus on the interaction of social and economic factors and the basic insights on policy that are
generated from that perspective.  This interaction raises its own set of questions of policy side effects, such as how they change social structure, that are also beyond the scope here (e.g., \cite{banerjeeetal2020})
but are important for future research.}
Third are subsidies to overcome basic financial poverty traps, including various loans, scholarships, subsidized day care and early childhood education, among others.
Fourth are policies aimed at overcoming homophily and opening up entrenched social networks.
These include enriching people's networks through enabling more contact across groups and
restructuring organizations; as well providing people with the information, opportunities, and role models that they are missing in their
existing networks---via affirmative action, mentorships, internships, and certification programs.
Importantly, network feedbacks and multiplier effects not only lead to large and persistent differences in behaviors and opportunities across groups, but also
help leverage well-designed and appropriately concentrated policies to be more effective in overcoming inequality and immobility.

Most all of the necessary policies have been used in various forms before.\footnote{However, many policies---such as
affirmative action---have been used at a small scale relative to what would be needed, for instance, to balance the demographics of those attending higher education with those of the general population (e.g., see \cite*{chetty2020income}).}
The
most novel policy discussed here is the use of small communities inside larger organizations and institutions to foster more cross-group exposure and interaction.
The idea of fostering teamwork and community bonds is part of many organizations, but this instead leverages what we have
seen in our network analysis to use small communities and teams build fruitful bridges across informational and opportunity barriers.

Importantly, a policy cocktail not only addresses different drivers and symptoms of inequality, but it can also be
designed to take advantage of complementarities between policies.
For example, enriching people's networks and information can have a greater effect if they have stronger
bases from early-childhood education and help in overcoming the investments and opportunity costs needed for higher education.
Conversely, offering subsidized loans for education to a disadvantaged group has a greater impact if they have good information about how to best make use of such resources in furthering
their education, and also if opportunities to become employed are eventually available.  Offering loans without information and opportunities will not have much impact, and similarly improving
information without the resources and eventual opportunities will not work, nor would making opportunities available to a population that does not
have the resources and information to take advantage of them.
All three of these policies are enhanced by the presence of the others.   This is a ``weakest-link'' challenge, where outcomes only change if a number of conditions are met and a failure of any of those
conditions leads to an overall failure.
Thus, policies that attack one condition at a time will be much less effective than ones that ensure that all conditions are met at the same time.
Further complementarities come from the fact that improvements in the outcomes for a member of some disadvantaged group puts that person in a better position to then offer information, opportunities, and serve as a role model for others within that group.  This is part of the network externalities.
Combining policies like affirmative action together with mentorships gives an extra boost to the power of each, by both improving the situation of some people in the network and then
enhancing the communication through the network.\footnote{Some policies are less costly and scale more easily, which is also an important consideration.
The point here is that combined policies, and ones that take into account social forces, can
be more effective that policies in isolation or ones that do not account for social structures.
Optimizing such policy cocktails to deal with costs is more context dependent. }  More generally, any policy that improves a person's situation then has network effects, and those can be further enhanced by
policies that help those improvements transfer through the network.
This applies not only to education and employment, but also to things like health, literacy, and other investments.

Taking this view means due to their network effects,
policies can have far-reaching and long-lasting impacts, as well as spillovers to other policies.  This makes it important to design combinations with a network perspective, and then also to evaluate them
by looking beyond the directly treated individuals.\footnote{For an approach to evaluating such policy cocktails, see \cite{banerjeeEtalVaccine2021}.}

There is much more that we need to understand, especially some of the potential side effects of any social engineering.
Networks serve many productive purposes and disrupting them can have unintended consequences.\footnote{There are other helpful policies that can have unintended consequences.  For instance,
as we find in \cite{banerjeeetal2020}, access to a microfinance program that can help people with investment and consumption smoothing can lead to the deterioration of social networks
even among people not involved in the program.}
Network effects can be highly nonlinear, and having a better understanding of how to concentrate policies within a network is an active area of research. More generally, we are involved in many networks of different forms of relationships: family, friends, co-workers, neighbors, etc., and they all provide different forms of interaction and support.   Understanding those distinctions, and the interactions
between different forms of networks, is another active area of research.

Finally, the rapid proliferation of technologies that mediate human interactions can have a profound impact on the shape of our networks.  The platforms that mediate many human interactions and the which information people see
are being designed by companies that are profit-motivated, and they are performing social engineering.  They provide the feeds and suggestions of connections, and often with deep knowledge of people's preferences and tendencies. While they are increasing
the volume and speed at which people interact, they can also be amplifying homophily and its impact.\footnote{For a look at some of the countervailing forces affecting assortative matching in the context of online dating sites see, e.g., \cite*{skopeksb2010,thomas2020online}.}
Moreover, platform-mediated interactions enable even greater network inequalities to emerge, as
some people amass not only thousands, but millions of followers.    Understanding how the incentives of platform designers deviate from social welfare, and
how to best leverage technology to improve our interactions and welfare is another important area for further study.

\begin{spacing}{0.88}
\bibliographystyle{jpe}
\bibliography{inequalityES}
\end{spacing}
\end{document}